\documentclass[12pt,preprint]{aastex}
\usepackage[margin=0.75in]{geometry}
\usepackage{graphicx}
\usepackage{natbib}
\usepackage{aas_macros}
\usepackage[section] {placeins}
\usepackage{subfigure}
\usepackage{float}
\usepackage{color}
\usepackage{footnote}

\usepackage{amsmath}

\def\msun{{\rm\,M_\odot}}

\def\rsun{{\rm\,R_\odot}} 
\def\msun{{\rm\,M_\odot}}

\newcommand{\kms}{\, {\rm km\, s}^{-1}}

\newcommand{\mpc}{\, {\rm Mpc}}
\newcommand{\kpc}{\, {\rm kpc}}
\newcommand{\hmpc}{\, h^{-1} \mpc}

\newcommand{\hi}{\mbox{H\,{\scriptsize I}\ }}
\newcommand{\heii}{\mbox{He\,{\scriptsize II}\ }}

\def\h2{${\rm\,H_2}$}

\makeatletter 

\def\hi{\noindent \hangindent=2.5em}

\def\kpc{{\rm\,kpc}}

\def\mpc{{\rm\,Mpc}}

\def\kms{{\rm\,km/s}}
\def\msun{{\rm\,M_\odot}}

\def\vol#1  {{{#1}{\rm,}\ }}

\def\hi{{H\thinspace I}\ }

\def\heii{{He\thinspace II}\ }

\newcount\refno
\newcount\rfno
\def\eq{$^{\the\refno\ }$\advance\refno by 1}
\def\ad{\advance\rfno by 1}

\def\clock{\count0=\time \divide\count0 by 60
     \count1=\count0 \multiply\count1 by -60 \advance\count1 by \time
     \number\count0:\ifnum\count1<10{0\number\count1}\else\number\count1\fi}

\def\myputfigure#1#2#3#4#5%
{\hskip0.03\textwidth\vskip#5pt
\makebox[0pt]{\hskip#2in
\includegraphics[width=#3\textwidth]{#1}}\vskip#4pt\hfill}

\newcount\refno
\newcount\rfno
\def\eq{$^{\the\refno\ }$\advance\refno by 1}
\def\ad{\advance\rfno by 1}

\definecolor{burntorange}{rgb}{1,0.4,0.2}
\DeclareMathOperator\erf{erf}

\newcommand{\lsim}{\raisebox{-0.13cm}{~\shortstack{$<$ \\[-0.07cm]
      $\sim$}}~}

\newcommand{\jlc}{j_{lc}}

\begin{document}

\title{On Post-Starburst Galaxies Dominating Tidal Disruption Events}

\author{
Renyue Cen$^{1}$ 
}

\footnotetext[1]{Princeton University Observatory, Princeton, NJ 08544; cen@astro.princeton.edu}

\begin{abstract} 

A starburst induced by a galaxy merger may create a relatively thin central stellar disk at radius $\le 100$pc.
We calculate the rate of tidal disruption events (TDEs) by the inspiraling secondary supermassive black (SMBH) through the disk.
With a small enough stellar velocity dispersion ($\sigma/v_c \le 0.1$) in the disk, it is shown that $10^5-10^6$ TDEs 
of solar-type main sequence stars per post-starburst galaxy (PSB) can be produced
to explain their dominance in producing observed TDEs.
Although the time it takes to bring the secondary SMBH to the disk apparently varies in the range of $\sim 0.1-1$Gyr since the starburst,
depending on its landing location and subsequently due to dynamical friction with stars exterior to the central stellar disk in question,
the vast majority of TDEs by the secondary SMBH in any individual PSB occurs 
within a space of time shorter than $\sim 30$Myr.
Five unique testable predictions of this model are suggested.

\end{abstract}

\section{Introduction}

When a star happens to plunge inside the tidal radius of a supermassive black hole, it will be torn apart, 
producing a tidal disruption event (TDE) that provides a useful tool to probe gas and stellar dynamics around SMBH,
and galaxy formation process potentially.
For a solar mass main sequence star, 
the tidal radius is greater than the Schwarzschild radius for a SMBH less massive than $10^8\msun$. 
For sub-giant or giant stars, still more massive SMBHs are also able to produce TDEs,
although their observable time scales become impractically long.

Post-starburst galaxies, sometimes called E+A or K+A's,
are characterized by spectra that are consistent with 
a starburst $0.1-1$Gyr ago followed by dormancy.
They constitute a fraction of $0.2-2\%$, depending on the observational definition, of all galaxies of comparable stellar masses
at low redshift \citep[e.g.,][]{2016Pattarakijwanich}.
Yet, current observations indicate 
that an overwhelming fraction of tidal disruption events (TDEs),
presumably normal main sequence stars tidally torn apart by SMBHs
at the center of galaxies, appear to occur in PSBs.
For example, all six TDEs observed by ASASSN survey
appear to occur in galaxies with spectral characteristics of PSBs \citep[][]{2016French,2017LawSmith,2018Graur}.
This suggests that stars in PSBs have a factor about $100$ more likely to provide TDEs.
For a recent survey of models, see an excellent review by \citet[][]{2018Stone}.
In this {\it Letter} a solution to this puzzle is sought and found.
We show that an inspiraling SMBH plowing through the stellar disk that is part of the starburst
can produce a sufficient number of TDEs to explain the observations.
We also suggest several tests for the model.

\section{Inspiral of Secondary SMBH Through a Nuclear Stellar Disk}

The physical setting of the problem in hand is as follows.
Two gas-rich galaxies each with a SMBH at their respective centers merge.
A starburst occurs in the process,
peaking at the time of the coalescence of the two galaxies,
followed by a rapid decline in star formation rate \citep[e.g.,][]{2006Hopkins}.
The merger of the two SMBHs may be delayed in time, relative to the starburst peak,
as simulations have shown.
The typical time delay is in the range of $0.1-1$Gyr,
not including an additional possible barrier at about parsec scale.
For the TDE rates derived in the present model, the parsec barrier has no effect.

We adopt a flat rotation curve throughout. 
High resolution cosmological zoom-in simulations covering galactic and central regions
with a resolution as high as 0.1pc \citep[][]{2010Hopkins, 2011Hopkins} support this assumption.
For the present purpose there is little to be gained by attempting to treat the situation with additional nuance than this.
While the stars dominate the gravity in this radial range, exterior to $r_{out}$ we assume the dark matter conspires to guarantee
a continuous flat rotation curve for simplicity and and we do not treat the region interior to $r_{in}$. 
We assume that the stellar subsystem is composed of 
a geometrically flat stellar disk with a mass fraction $\eta$ and 
a spherical component with a mass fraction $1-\eta$.
In the radial range of $[r_{in},r_{out}]$,
the stellar volume mass density in the disk can be expressed as
\begin{equation}
\label{eq:rho}
  \begin{split}
\rho_*(r) = {\eta v_c^2\over 4\pi G r^2}({v_c\over\sigma}),
  \end{split}
\end{equation}
\noindent
where $v_c$ and $\sigma$ are the rotation velocity of and velocity dispersion (assumed to be isotropic) in the disk, 
respectively, at the cylindrical radius, $r$.
The Mestel stellar disk's mass surface density is 
\begin{equation}
\label{eq:Sigmastar}
  \begin{split}
\Sigma_*(r) = {\eta v_c^2\over 2\pi G r}.
  \end{split}
\end{equation}
\noindent
Let us for simplicity assume a single population of solar mass stars to yield the stellar number density in the disk
\begin{equation}
\label{eq:nstar}
  \begin{split}
n_*(r) = {\eta v_c^3\over 4\pi G r^2 \sigma\msun}.
  \end{split}
\end{equation}
\noindent
Given this physical backdrop, the process that we are interested in is 
the inspiral of the secondary SMBH through the flat stellar disk.
We denote the inspiraling SMBH as ``the secondary" of mass $M_2$, as opposed to the central SMBH denoted of mass $M_1$.
To present a concrete set of quantitative results we shall choose a fiducial case of two merging galaxies each 
with a SMBH of mass $M_1=M_2=10^{7}\msun$ and $v_c=159\kms$ to denominate relevant terms, 
following the relation between galaxy stellar mass and SMBH mass.
The merged galaxy is assumed to slide along
the \citet[][]{2002Tremaine} relation so to have a rotation velocity of $v_c=2^{1/4}\times 159\kms = 189\kms$,
which is assumed to have achieved after the merger of the galaxies but prior to inspiral of the secondary through the central stellar disk.
The total stellar mass interior to $r$ (including both the disk and bulge):
\begin{equation}
\label{eq:Mr}
  \begin{split}
M(<r) = 10^7 ({r\over 1.2{\rm pc}}) ({v_c\over 189\kms})^2 \msun.
  \end{split}
\end{equation}
\noindent
On the grounds that dynamical friction induced inspiral stalls at the radius where
the interior stellar mass on the disk is equal to the mass of the inspiraling SMBH,
we define the inner radius $r_{in}$ as 
\begin{equation}
\label{eq:rin}
  \begin{split}
r_{in} = 1.2 ({M_2\over 10^7 \msun})  ({v_c\over 189\kms})^{-2}~{\rm pc}.
  \end{split}
\end{equation}
\noindent

When the secondary, if with zero orbital eccentricity, moves at the circular velocity $v_c$ at any given radius,
the stars at the same radius moves at a lower azimuthal velocity $v_\phi$.
The asymmetric drift, 
$v_{\phi}-v_c$,
the relative velocity of stars to a notional circular velocity at the radius,
is governed physically by the Jeans third equation \citep[][]{1987Binney}
and observed in our solar neighborhood \citep[e.g.,][]{2013Golubov, 2014Sharma}.
For an isotropic velocity dispersion of stars in the disk with a local dispersion
$\sigma\equiv\sigma_R=\sigma_\phi=\sigma_z\ll v_c$, which we shall assume, and for a flat rotation curve,
we have 
\begin{equation}
\label{eq:v2}
  \begin{split}
v_2\equiv v_c - v_{\phi}={\sigma^2\over v_c}=0.1\sigma ({10\sigma\over v_c}),
  \end{split}
\end{equation}
\noindent
For a relatively thin disk with $\sigma\le 0.1 v_c$ that are of relevance here, $v_2 \ll \sigma$.
Note that $v_{\phi}<v_c$, i.e., stars collectively move more slowly than the circular velocity at that location.
The physical meaning of the asymmetric drift is easily understood in terms of the presence of an equivalent negative radial pressure gradient 
in the stars due to local velocity dispersion, as the Jeans equation displays.
This lag, direction-wise, may be understood in another intuitive way.
Stars with non-zero velocity dispersion, i.e., not strictly on circular orbits,
have non-zero eccentricities.
In any non-Keplerian orbit, which is the case here for a flat rotation curve,
the epicyclic frequency is larger than the azimuthal frequency,
causing perigalacticon to precess backwards relative to zero eccentricity orbits.

Because of finite $v_2$, the secondary experiences a dynamical friction force.
This is important because it means that the secondary in a circular orbit in a disk in the absence of any bulge component 
can still experience a dynamical friction and move inward.
In a two-dimensional configuration the primary dynamical effect is due to close encounters between the secondary and stars
\citep[][]{1972Rybicki}, as opposed to the usual three-dimensional configuration where distant encounters dominate 
\citep[][]{1943Chandrasekhar}.
If the influence radius of the secondary, defined as $r_2\equiv GM_2/\sigma^2$, is greater than the half-thickness of the stellar disk, $h$,
then the situation is considered to be two-dimensional. We have 
\begin{equation}
\label{eq:r2h}
  \begin{split}
{r_2\over h} = 1.2 ({10\sigma\over v_c})^{-3} ({M_2\over 10^7\msun})({v_c\over 189\kms})^{-2} ({r\over 1kpc})^{-1}.
  \end{split}
\end{equation}
\noindent
Thus, for the fiducial case considered of $M_2=10^7$ and $v_c=189\kms$, 
in the regime of interest here with $\sigma/v_c\le 0.1$, the two-dimensional condition is satisfied for radius $r\le 1$kpc.
The two-dimensional dynamical friction force is \citep[][Eq III.3]{1986Quinn} 
\begin{equation}
\label{eq:Fdf}
  \begin{split} 
F_{df} = -2\pi G \Sigma_* M_2\left\{{\sqrt{2\pi}\over 4}{v\over \sigma}\exp{(-{v^2\over 4\sigma^2})}\times\left[I_0({v^2\over4\sigma^2})+I_1({v^2\over 4\sigma^2})\right]\right\},
  \end{split}
\end{equation}
\noindent
where $v$ is the velocity of the secondary relative to the stars,
$I_0$ and $I_1$ are modified Bessel functions of the first kind \citep[][]{1972Abramowitz}.
In the limit $v \ll \sigma$, 
\begin{equation}
\label{eq:Fdf2}
  \begin{split} 
F_{df} = -\sqrt{\pi^3\over 2} {G \Sigma_* M_2v\over \sigma}.
  \end{split}
\end{equation}
\noindent
In this limit, for our case, a rotating disk,  
we may follow the procedure of \citet[][]{1943Chandrasekhar} by elementarily integrating the spatial range on the disk from $-h$ to $+h$
over which the shear velocity is subdominant to the velocity dispersion 
(along with the integrations over the distribution over the angle between velocity vectors and the Maxwellian velocity distribution)
to derive the frictional force:
\begin{equation}
\label{eq:Fdf3}
  \begin{split} 
F_{df}^\prime = -{3\sqrt{2\pi} h \Sigma_* \sigma v} = - {3\sqrt{2\pi} G \Sigma_* M_2 v\over \sigma}.
  \end{split}
\end{equation}
\noindent
It is seen that Eq (\ref{eq:Fdf3}) and Eq (\ref{eq:Fdf2}) differ only by a factor
of unity ($\pi/6$), reflecting again close encounters being largely responsible for dynamical frictional force in 
the two-dimensional case.
For simplicity, without introducing a large error, and given the ambiguity in choosing the radial extent 
of integration used to derive Eq (\ref{eq:Fdf3}), we just use Eq (\ref{eq:Fdf2}) for all subsequent calculations.
The dynamical friction time for the two-dimensional component is then
\begin{equation}
\label{eq:tdf2dp}
  \begin{split}
t_{2d}\equiv ({d\ln r\over dt})^{-1} = {M_2 v_c \over F_{df}} {v\over v_2} = {r v\over \eta v_c v_2} \left\{{\sqrt{2\pi}\over 4}{v\over \sigma}\exp{(-{v^2\over 4\sigma^2})}\times\left[I_0({v^2\over4\sigma^2})+I_1({v^2\over 4\sigma^2})\right]\right\}^{-1}, 
  \end{split}
\end{equation}
\noindent
where the $v_2/v$ factor is the tangential fraction of the dynamical friction force,
and $v$ is the total velocity of the secondary relative to local stars,
\begin{equation}
\label{eq:v}
  \begin{split}
v =\sqrt{v_2^2 + v_r^2}\quad\quad {\rm with} \quad\quad v_r = {r\over t_{df}}
  \end{split}
\end{equation}
\noindent
being the radial drift velocity of the secondary and $v_2$ the asymmetric drift velocity 
(Eq \ref{eq:v2}).
In addition, the dynamical friction time due to the three-dimensional component is
\begin{equation}
\label{eq:t3d}
t_{3d} = {2v^3 r^2 [\erf(X)-2X\exp{(-X^2)}/\sqrt{\pi}]^{-1} \over 3 v_c^2 \ln\Lambda G M_2  (1-\eta)}
\end{equation}
\noindent
\citep[][]{1987Binney}, where $X=v/\sqrt{2}\sigma$ and we adopt a Coulomb logarithm $\ln\Lambda$ equal to three.
Then, the overall dynamical friction time is 
\begin{equation}
\label{eq:tdf}
t_{df} = (t_{2d}^{-1} + t_{3d}^{-1})^{-1},
\end{equation}
\noindent
which will be used throughout our subsequent calculations.
\begin{figure}[!h]
\centering
\vskip -0.0cm
\resizebox{6.0in}{!}{\includegraphics[angle=0]{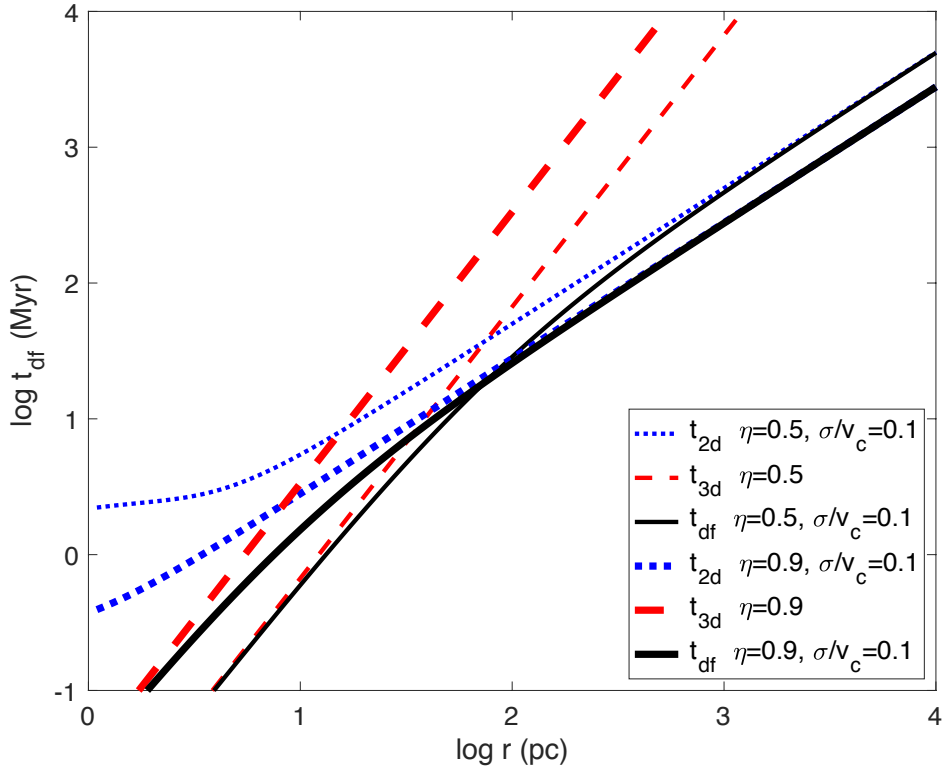}}
\vskip 0.0cm
\caption{
shows the dynamical friction time (${\rm t_{df}}$, Eq \ref{eq:tdf}) for two cases:
$\eta=0.5$ with $\sigma/v_c=0.1$ (solid thin black curve) and
$\eta=0.9$ with $\sigma/v_c=0.1$ (solid thick black curve),
both with $v_c=189\kms$ and $M_2=10^7\msun$.
Also shown as the dotted blue and dashed red curves 
are their respective two-dimensional (${\rm t_{2d}}$, Eq \ref{eq:tdf2dp}) 
and three-dimensional dynamic friction time (${\rm t_{3d}}$, Eq \ref{eq:t3d}).
In the two-dimensional case, the flat regime at the small radius end 
is due to the dynamical friction time that is constrained by the limited radial range  due to 
the rotational velocity shear (${\rm t_{df, 2d}}$, Eq \ref{eq:tdf2dp}),
whereas the ascending portion at the large radius end is determined by 
$t_{df, 2d}$ (Eq \ref{eq:tdf2dp}).
}
\label{fig:tdf}
\end{figure}
Figure~\ref{fig:tdf} shows the dynamical friction time (${\rm t_{df}}$, Eq \ref{eq:tdf}) for two cases:
$\eta=0.5$ with $\sigma/v_c=0.1$ (solid thin black curve) and
$\eta=0.9$ with $\sigma/v_c=0.1$ (solid thick black curve), both with $v_c=189\kms$ and $M_2=10^7\msun$,
along with the breakdowns due to the two-dimensional and three-dimensional components.
We see that for $\eta<0.9$ and $\sigma/v_c=0.1$
the overall dynamical friction induced inspiral is due to the three-dimensional component at $r\le 10$pc;
in fact, for any applicable cases (see figures below),
in the inner region of $r=1-10$pc 
the three-dimensional dynamical friction dominates and sets the time scale of the inspiral 
in that radial range.
It is important to note, however, that the TDE rate is mainly due to the interaction of the inspiraling secondary and stars in the disk,
as we show below, thanks to its high volume density of stars. 
Let us now examine the TDE rate by the secondary during its inspiral.
Because the stars are essentially collisionless, they can accrete onto the secondary only at a rate about equal to
the cross section of the secondary times the mass flux, $v \rho_*$ \citep[][]{1926Eddington}, as opposed to a higher, Bondi rate for collisional matter.
The effective cross section of the secondary may be identified with the tidal capture cross section, which is larger than but on the same order as
the tidal disruption cross section, although what happens to the stars once captured is complex.
We estimate the TDE rate based on stars that directly plunge into the radius twice the tidal radius.
We now derive a general expression of TDE events for both non-zero relative bulk velocity of the secondary to stars
and non-zero velocity dispersion of stars with the latter being assumed to already have a relaxed Maxwellian distribution.
If the secondary moves through a static sea of stars of density $n_*$ at a velocity $v$, the rate of stars entering the loss cone would be
\begin{equation}
\label{eq:Rv}
  \begin{split} 
R(v|\sigma=0) = \pi r_t^2 n_* v (1 + {2GM_2\over v^2 r_t}),
  \end{split}
\end{equation}
\noindent
where $r_t$ is the tidal radius of the loss cone surface:
\begin{equation}
\label{eq:rt}
  \begin{split}
r_t = r_* ({M_2\over m_*})^{1/3} & = 1.5\times 10^{13} ({M_2\over 10^7\msun})^{1/3} ({r_*\over \rsun}) ({m_*\over \msun})^{-1/3}~cm \\
&=1.09 r_{sch}(M_2) ({M_2\over 10^8\msun})^{-2/3} ({r_*\over \rsun}) ({m_*\over \msun})^{-1/3},
  \end{split}
\end{equation}
\noindent
with $m_*$ and $r_*$ being the stellar mass and radius, respectively, 
and $r_{sch}(M_2)$ the Schwarzschild radius of the secondary.

For the secondary moving through stars with a Maxwellian velocity distribution of dispersion $\sigma$ at a mean relative velocity $v$,
one may convolve $R(v|\sigma=0)$ in Eq (\ref{eq:Rv}) with the velocity distribution to obtain the overall rate.
Choosing the direction of $v$ in plus x-direction, we have
\begin{equation}
\label{eq:Rv2}
  \begin{split} 
R(v,\sigma) = &\int_{-\infty}^{+\infty}\int_{0}^{+\infty}\pi r_t^2 n_* \sqrt{(v-v_x)^2 + v_t^2} \left\{1 + {2GM_2\over [(v-v_x)^2+v_t^2] r_t}\right\} \\
 & \times {1\over \sqrt{2\pi}\sigma^3} \exp{\left[-(v_x^2+v_t^2)/2\sigma^2\right]} v_t dv_x dv_t, 
  \end{split}
\end{equation}
\noindent
where $v_t^2 = v_y^2 + v_z^2$,
and the outer and inner integrals are for $v_x$ and $v_t$, respectively.
With a bit manipulation one finds 
\begin{equation}
\label{eq:Rv3}
  \begin{split} 
R(v,\sigma) &= \pi r_t^2 n_* \sigma \left\{\sqrt{2\over \pi} \exp{(-v^2/2\sigma^2)} + {v\over\sigma} \erf{({v\over\sqrt{2}\sigma})} + {2\sigma\over v} \left[\exp{(-v^2/2\sigma^2)} -1 + \erf{({v\over \sqrt{2}\sigma})}\right] \right\} \\
 & + {\pi n_* j_{lc}^2 \over v} \left[\erf{({v\over \sqrt{2}\sigma})} + 1 - \exp{(-v^2/2\sigma^2)}\right], 
  \end{split}
\end{equation}
\noindent
where the first and second terms correspond to their counterparts in Eq (\ref{eq:Rv}) with the latter due to gravitational focusing,
and $j_{lc}$ is the angular momentum at loss cone surface about the secondary on a circular orbit:
\begin{equation}
\label{eq:jlc}
  \begin{split}
\jlc=\sqrt{GM_2^{4/3}m_*^{-1/3}r_*} = 1.4\times 10^{23} ({M_{2}\over 10^7\msun})^{2/3} ({m_*\over \msun})^{-1/6} ({r_*\over \rsun})^{1/2} cm^2/s.
  \end{split}
\end{equation}
\noindent
For extreme events like TDEs the orbital velocity at tidal radius $r_t$ is much larger than typical velocity
of stars at infinity relative to the secondary, so the second term in 
Eq (\ref{eq:Rv3}) dominates.
Thus, for the sake of conciseness we shall neglect the first term with negligible loss of accuracy in our case.
The radial distribution of TDEs may be expressed as
\begin{equation}
\label{eq:dNdr}
  \begin{split}
{dN_{tde}\over d\ln r} & = R(v,\sigma) t_{df} = {\eta v_c^3 j_{lc}^2 t_{df}\over 4 G m_* \sigma v r^2} \left[\erf{({v\over \sqrt{2}\sigma})} + 1 - \exp{(-v^2/2\sigma^2)}\right]. 
  \end{split}
\end{equation}
\noindent

A key notable point 
in terms of the time scale is that
the vast majority of TDEs in a PSB in our model likely occur within a time scale
that is significantly less than the age since starburst of PSBs of $0.1-1$Gyr.
Thus, if our model were to explain the observed TDEs in PSBs,
which show  an apparent delay, relative to the starburst event itself, of up to $\sim 1~$Gyr,
this indicates that it is the time that it takes to bring the secondary
into the central stellar disk region and to be co-planar that determines the observed temporal distribution relative to the starburst,
before the secondary interacts with the central stellar disk
that subsequently dominates the TDE events.
Such an expectation is quite plausible in the context of galaxy mergers, as evidenced by galaxy merger simulations.

A systematic simulation survey of black hole mergers in the context of galaxy mergers is not available,
due to the computational cost,  daunting physical complexity and a large parameter space.
Nonetheless, valuable information from existing simulations may be extracted.
Our survey of literature is by no means exhaustive but hoped to be representative.
In the merger simulations of \citet[][]{2006Hopkins} it is seen in their Figure 13 that
the final starburst occurs at $1.5$Gyr since the beginning of the merger for a black hole pair of mass $3\times 10^7\msun$ each.
We can not find information about the black hole separation at this time.
But from their visualization plots it seems that by this time the galaxies are largely merged,
with separations likely less than a few kpc at most.
In \citet[][Figure 14]{2009bJohansson} one sees 
that by the time the starburst ends at simulation time $t\sim 1.8$Gyr,
the separation of the binary BHs is $\sim 1$kpc.
Using the three-dimensional dynamical friction time formula \citep[][]{1943Chandrasekhar},
we find $t_{DF}=1.7$Gyr and $0.43$Gyr for a $1\times 10^7\msun$ black hole at $1$kpc and $0.5$kpc, respectively,
in a spherical system with a circular velocity of $189\kms$. 
In the 1:4 merger simulations 
\citet[][]{2009Callegari} find that once the separation of the galaxy pair (and BH pair) reaches $10$kpc,
it takes about $0.5$Gyr to reach $\sim 0.1$kpc.
This suggests that so long as the starburst does not end before the BH reaches $10$kpc separation,
the BH merger would occur in the time frame of $0.1-1$Gyr.
In the most comprehensive study so far
\citet[][]{2017Tamburello} find that the black hole pair reaches a separation of $\sim 100$pc 
in the range of $0.11-0.79$Gyr from a sample of about two dozen merger simulations (see Table 2 in their paper),
although there is a small fraction of cases where mergers never occur.
Observationally, \citet[][]{2017French} infer a post-starburst age in the range 
$0.06-1$Gyr from eight TDE cases; 
when $1\sigma$ error bars are included, the range of post-starburst ages extends to $0.05-1.2$Gyr.
This range of PSB age of $\le 1Gyr$ seems accommodatable by the galaxy merger dynamics 
to bring the secondary close to the central region from extant simulations.
As it is clear now that 
it is the total number of TDEs per galaxy that is predicted for a given 
physical configuration of the system, including $\sigma/v_c$, $\eta$, $v_c$ and $M_2$.
If, for some reason, the secondary black hole reaches the central disk
in a shorter span of time since the starburst for some subset of starburst galaxies,
then their apparent rate will be inversely proportional to time interval between the starburst to the arrival at the central disk.
Perhaps the apparently higher rate of TDEs in ULIRGs \citep[][]{2017Tadhunter} is due to this reason.

One important requirement 
concerns bringing the secondary to be co-planar with the central stellar disk.
Mergers of two galaxies possess some axisymmetry dictated by the orbital angular momentum of the merger and formation of a disky component
due to gas dissipational processes.
Thus, it is likely that the orientation of the orbit of the secondary may be largely co-planar initially.
\citet[][]{2017Tamburello} show that a flat disk is formed in the central region due to gas inflow,
although the exact scale height is likely limited by their finite resolution.
Without rigorous proof one has to contend with the possibility that the secondary 
is not exactly co-planar with the central stellar disk, when it is still at some large radius.
Even in this case, the orbital plane of the secondary will be re-aligned with the central stellar disk during the inward migration
via dynamical friction.
\citet[][]{1977Binney} shows that
in an oblate system with anisotropic velocity distribution,
the dynamical friction drag tends to align the inspiraling object with the disk plane, so long as not on a polar orbit initially.
The timescale on which this occurs is precisely the timescale for the action of dynamical friction.
The basic analytic framework of \citet[][]{1977Binney} is shown to provide
a much better agreement with simulations for inclination dependent dynamical friction time scale than
the classic formulation of \citet[][]{1943Chandrasekhar} for flattened systems.
More importantly, the decay rate of the orbital inclination that is not observed using the classic approach 
is quantitatively reproduced in simulations \citep[][]{2004Penarrubia} when anisotropic dynamical friction 
formulae \citep[][]{1977Binney} are used.
In the simulations \citep[][]{2002Penarrubia, 2004Penarrubia} a relatively modest amount of anisotropy ($q=0.6$) is employed
for the dark matter halo to show the efficacy of the inclination decay of satellite orbits in a flattened host system.
In the inner regions of interest here, baryons dominate dynamically and starburst is presumably triggered by
a strong gas inflow due to galaxy merger, and turbulent dissipation and gas cooling are likely strong to yield flattened systems.
This is of course fully in accord and self-consistent with the presumed existence of a thin flat central stellar disk that is 
the foundation of our working hypothesis. 
The inclination decay of the secondary, if initially exists, 
can be due to dynamical friction with the stars on a larger spatial scale with an overall anisotropic velocity distribution,
i.e., larger than the central stellar disk of size of $\sim 100$pc,
that operate on a time scale likely in the range of $0.1-1$Gyr.
Note for example the dynamical friction time is 
we find $t_{DF}=1$Gyr and $0.1$Gyr for a $1\times 10^7\msun$ black hole at $0.77$kpc and $0.24$kpc, respectively,
in a spherical system with a circular velocity of $189\kms$. 
Thus, the co-planar condition for the orbital plane of the secondary and the central stellar disk is physically plausible,
when it reaches the outer edge of the central disk. 
Even if the central disk and the orbit of the secondary is misaligned when the latter reaches the outer edge of the former,
dynamical friction from that point onward will subsequently align it with the disk on the dynamical friction time scale, i.e., order of an e-folding in radius. 
Since most of the TDEs occur in the innermost region, one or two e-folding in radius can be spent to re-align 
the secondary with the central stellar disk with little effect on the overall TDE rate (and repeating time scale).

Another issue worth clarifying is the orbital eccentricity of the inspiraling black hole, since
we have implicitly assumed zero eccentricity in the derivation of 
Eq (\ref{eq:Fdf},\ref{eq:Fdf2},\ref{eq:Fdf3}).
However, this assumption serves only as a sufficient but not necessary condition for dynamical friction to operate.
That it, even in a zero eccentricity orbit, the secondary still experiences dynamical friction force due to 
the non-zero asymmetric drift velocity $v_2=\sigma^2/v_c$.
Any significant eccentricity would render the relative velocity of the inspiraling black hole to the embedding stars 
possibly significantly above $\sim \sigma^2/v_c$,
which would increase an additional dynamic friction force in the radial direction,
leaving the tangential dynamic friction force unchanged.
Nonetheless, one notes that if the secondary were in a radial orbit, then the ``cruise" radial velocity due to the balance between gravity and the dynamical friction 
force due to the three-dimensional component (that dominates at small radii) can be shown to be equal to $v_c$.
In this case, we find that the total number of TDEs per PSB is in the range of $\le 10^3$ for the fiducial case of $M_2=10^7$ and
$v_c=189\kms$. Such a case would be much lower and hence inconsistent with the observationally inferred TDE rate
of $10^{5}-10^{6}$ per PSB \citep[][]{2016French}.
Therefore, one needs to make sure that increasing radialization of the orbit of the secondary is avoided,
if the initial eccentricity is not identically zero.

We now check two approximately bracketing cases to settle the issue.
First, let us continue to consider the case of an isothermal sphere density profile.
The apsides in a gravitational potential $\phi(r)$
with specific energy $E$ and specific angular momentum $J$
are the two roots of the following equation \citep[Eq 3-13 in][]{1987Binney}:
Defining the orbital eccentricity $e$ as $r_a/r_p=(1+e)/(1-e)$ with $r_a=r_0(1+e)$ [and $r_p=r_0(1-e)$],
where $r_p$ and $r_a$ are the perigalacticon and apogalacticon distance, respectively,
it can then be shown that, to the lowest order in $e$, the specific total energy and specific angular momentum are 
\begin{equation}
\label{eq:E}
  \begin{split}
E = ({1\over 2} + {7 e^2\over 6}) v_c^2 \quad\quad and \quad\quad J = \sqrt{1 - {5 e^2\over 3}} v_c r_0,
  \end{split}
\end{equation}
\noindent
where we have defined the normalization of the logarithmic gravitational potential energy for an isothernal density profile such that
${\rm \phi(r) = v_c^2\ln{r\over r_0}}$ without loss of generality.
Note that additional, higher order terms in $e$ would be needed when $e\to \sqrt{3/5}$ as Eq (\ref{eq:E}) shows
and it is also possible that orbits become unstable when $e$ becomes too large.
We consider here that $e$ is not too large initially.
From Eq (\ref{eq:E}) it is seen that 
$E$ is a function of and decreases with decreasing eccentricity $e$.
This indicates that in the presence of any energy dissipation, the orbit
tends to zero eccentricity.
It is also seen that the rate of decrease of eccentricity is $de/dE \propto {1/e}$ hence the time scale 
of circularization takes place on the similar time scale as the energy dissipation time scale (i.e, 
the dynamical friction time scale) when $e\gg 0$ but accelerates when $e\to 0$.
Thus, the circularization time scale is about equal to dynamical friction time scale,
if the orbit starts with a significant eccentricity but may take a much shorter time scale
for an initially nearly circular orbit.
We stress that this outcome of circularization is derived based on
a logarithmic potential corresponding to a flat rotation curve. 
While it is a good assumption, as simulations have shown,
it is still prudent to stress that circularization is not necessarily the only outcome in general, as we show now.

Consider next the following simplified problem: 
the black hole moving in an eccentric orbit about a dominant point mass  
is subject to a frictional force that is a function of both the distance to the center and velocity.  
We assume that the gravitational effect due to the frictional matter is negligible.
A further simplication is made for the convenience of calculation: the dynamical effect due to the frictional force
is small enough so that a Keplerian (closed) orbit remains a good approximation for each full orbit.
We adopt the units such that the specific total energy of the orbiting black hole is ${\rm E = -{1\over 2}}$,
and the specific angular momentum is
$J = \sqrt{1-e^2}$.
With the familiar expressions
for the distance to the focus $r$, the tangential velocity $v_\phi$ 
and the magnitude of the total velocity $v$:
\begin{equation}
\label{eq:radius}
r = {(1-e^2)\over (1+e\cos\theta)}, \quad\quad v_{\theta} \equiv {rd\theta\over dt} = {(1+e\cos\theta)\over (1-e^2)^{1/2}}, \quad\quad v = \sqrt{{2\over r}-1},
\end{equation}
\noindent
where $\theta$ is the true anomaly, being zero at perigalacticon. 
Since ${\rm e^2 = 1 + 2E J^2}$, utilizing various expressions above, 
we have
\begin{equation}
\label{eq:de}
\Delta e = (1-e^2)^{1/2}e^{-1}\left[(1-e^2)^{1/2}\Delta E - \Delta J\right] \equiv (1-e^2)^{1/2}e^{-1}I,
\end{equation}
\noindent
where we shall define $\Delta e$, $\Delta E$ and $\Delta J$ as the change of eccentricity,
specific total energy and specific angular momentum, respectively, per full radial orbit.
We now examine the term $I$ defined by the last definition equality in Eq (\ref{eq:de}).
To be tractable, let the acceleration due to frictional force have the following powerlaw velocity and  
radial dependencies: 
\begin{equation}
\label{eq:a}
\vec{\mathrm{a}} = -A r^\alpha v^\beta \vec{\mathbf{v}},
\end{equation}
\noindent
where $A$ is a positive constant, $\beta$ a constant, $\alpha$ a constant slope,
and $\vec{\mathbf{v}}$ and $v$ the velocity vector and its magnitude,
noting that the radial dependence is inherited from the density's radial profile, ${\rm \rho(r) \propto r^{\alpha}}$.
While $\alpha$ may be non-positive in most physical contexts,
our derivation does not impose any constraint. 
Gathering, we express 
\begin{equation}
\begin{split}
\label{eq:I}
I &\equiv (1-e^2)^{1/2}\Delta E - \Delta J \\
&= -2(1-e^2)^{1/2}A\int_0^{P/2}r^\alpha v^\beta  (\vec{\mathbf{v}}\cdot \vec{\mathbf{v}}) {dt + 2A\int_0^{P/2} r^\alpha v^\beta} |\vec{\mathbf{r}}\times \vec{\mathbf{v}}|dt\\
&= 2A\int_0^{\Theta} r^{\alpha-\beta/2+1}(2-r)^{\beta/2}(r-1)d\theta,
\end{split}
\end{equation}
\noindent
where $P$ is the period of a full radial orbit with the integration going from perigalacticon to apogalacticon,
and $\Theta$ the azimuthal advance per half radial period, equal to $\pi$ in this case of closed orbits.
To proceed,
we change the integration element from $d\theta$ to the length element along the ellipse $dl=r d\theta$. 
Now the last equality in Eq (\ref{eq:I}) becomes 
\begin{equation}
\begin{split}
\label{eq:I2}
I &= 2A\int_0^{C/2} r^{\alpha-\beta/2}(2-r)^{\beta/2}(r-1)dl,
\end{split}
\end{equation}
\noindent
where $C/2$ is the half circumference of the orbit with the integration going from perigalacticon to apogalacticon.
With the integration variable now changed to $l$ that is invariant of the vantage point, 
one is free to move the center from
one focus to the other by switching the radius from $r$ to $2-r$ to obtain an identity
\begin{equation}
\begin{split}
\label{eq:I3}
I = -2A\int_0^{C/2}r^{\beta/2}(2-r)^{\alpha-\beta/2}(r-1)dl.
\end{split}
\end{equation}
\noindent
Taking the arithmetic average of 
Eq (\ref{eq:I2}) 
and
Eq (\ref{eq:I3}), one obtains 
\begin{equation}
\begin{split}
\label{eq:I4}
I = \int_0^{C/2} A(r-1)[r^{\alpha-\beta/2}(2-r)^{\beta/2} - r^{\beta/2}(2-r)^{\alpha-\beta/2}]dl.
\end{split}
\end{equation}
\noindent
One sees that for $\alpha=\beta$,
$I$ in Eq (\ref{eq:I4}) is identically zero,
meaning $\Delta e=0$ in Eq (\ref{eq:de}) for any initial $e$.
This thus indicates that the orbital eccentricity of a slowly inspiraling black hole under 
a frictional force of the form $-A(rv)^\alpha\vec{\mathbf{v}}$ (where $A$ is a positive constant) is non-changing.
For $\alpha>\beta$, $\Delta e$ will be greater than zero, meaning that the orbit
will be increasingly radialized during the inspiral, 
whereas for $\alpha<\beta$  the orbit will be increasingly circularized.
Physically, this can be understood as 
a result of relatively higher loss of angular momentum per unit loss of energy hence gain of eccentricity at the perigalacticon 
as compared to a lower gain of angular momentum per unit gain of energy hence gain of eccentricity at the apogalacticon, for $\alpha>\beta$,
thus leading to a net radialization over a complete orbit.
For $\alpha<\beta$, the opposite holds.
Let us consider two relevant applications of this result.
First, in the standard three-dimensional dynamical friction case \citep[][]{1943Chandrasekhar},
$\beta=-3$. Thus, unless the density slope is steeper than $-3$,
the eccentricity is to increase under such frictional force, thus leading to radialization of the orbit spiraling inward.
Second, in the standard two-dimensional dynamical friction case (Eq \ref{eq:Fdf}), we $\beta=-1$ for $v/\sigma\gg 1$ and $\beta=0$ for $v/\sigma\ll 1$. 
Therefore, in this case, for any density profile that increases with decreasing radius, the orbit tends to circularize with time.
In a more detailed calculation using dynamical friction formula that includes effects due to stars moving faster than
the inspiraling black hole, \citet[][]{2017Dosopoulou}
conclude that for $\alpha \lsim -2$, the orbit of the inspiraling black hole tends to radialization,
overlapping with the radialization range of $\alpha \le -3$ found here.

Overall, considerations of two bracketing examples suggest that, on the one hand, 
orbital circularization is likely achieved if the density profile is close isothermal 
regardless whether the medium for dynamical friction is also gravitationally dominant.
On the other hand, at the other end of the spectrum where the central mass gravitationally dominates,
the dynamical friction may lead to radialization if the velocity distribution of the medium is largely three-dimensional,
whereas it leads to circularization if the velocity distribution of the medium is largely two-dimensional.
Since in an oblate velocity distribution, dynamical friction leads to inspiraling black hole becoming co-planar,
circularization should also ensue in this case so long as enough dynamical friction takes place after becoming co-planar.
Thus, in the physical configuration of an overall oblate stellar distribution along with a thin stellar disk
in the central region that we propose here, the only likely situation where circularization does not occur is when the secondary black hole
directly lands at a radius to which the interior stellar mass is not significantly greater than the mass of the inspiraling black hole.
Such a situation is not expected to happen in practice.
Anyway, since we have already assumed that the inner $r_{in}$ 
(Eq \ref{eq:rin}) is where the interior stellar mass of the disk is equal to the mass
of the inspiraling black hole, such a situation is moot.

\section{Predictions}\label{sec: pred}

\subsection{TDE Repeaters}

To illustrate, in the limit $v\gg \sigma$, which is the case when the secondary has migrated 
into the inner region of the disk,
the second term of Eq (\ref{eq:Rv3}) gives 
\begin{equation}
\label{eq:t}
  \begin{split}
R(v,\sigma) = {\eta v_c^3 j_{lc}^2 \over 2G m_* \sigma v r^2} = 0.034 \eta ({v_c\over 189\kms}) ({v_c\over 10\sigma}) ({v_c\over v}) ({j_{lc}\over 1.2\times 10^{23} {\rm cm^2/s}})^2({r\over 1{\rm pc}})^{-2}~{\rm yr}^{-1},
  \end{split}
\end{equation}
\noindent
which is indicative that TDEs may re-occur in the same PSBs within an accessible time scale.
To gain a more quantitative assessment, we have performed a simple analysis with the following steps.
(1) We use (the inverse of) Eq (\ref{eq:Rv3}) to obtain the mean expectation value of time interval between two successive TDEs, $\bar{{\Delta t}}$,
when one just occurred at a radius $r$.
(2) With the expectation value $\bar{{\Delta t}}$ we use the normalized Poisson distribution to obtain the probability distribution function
as a function of time interval ($\Delta t$) between the TDE that just occurred at $r$ and the next one, $P(r,\Delta t)$.
(3) We convolve $P(r,\Delta t)$ with Eq (\ref{eq:dNdr}) to obtain 
the overall mean probability distribution function as a function of time interval, $P(\Delta t)$.
Generally, $P(\Delta t)$ is a function of three variables,
$\eta$, $\sigma/v_c$ and $M_2$ (if $M_1$ can be related to $M_2$ or expressed by $v_c$).
The total number of TDEs per PSB, $N_{tde}$ (Eq \ref{eq:dNdr}), is a function of 
$\eta$, $\sigma/v_c$ and $M_2$ as well. Therefore, if observations can provide constraints on $N_{tde}$,
only two degrees of freedom are left.

\begin{figure}[!h]
\centering
\vskip -0.0cm
\resizebox{6.0in}{!}{\includegraphics[angle=0]{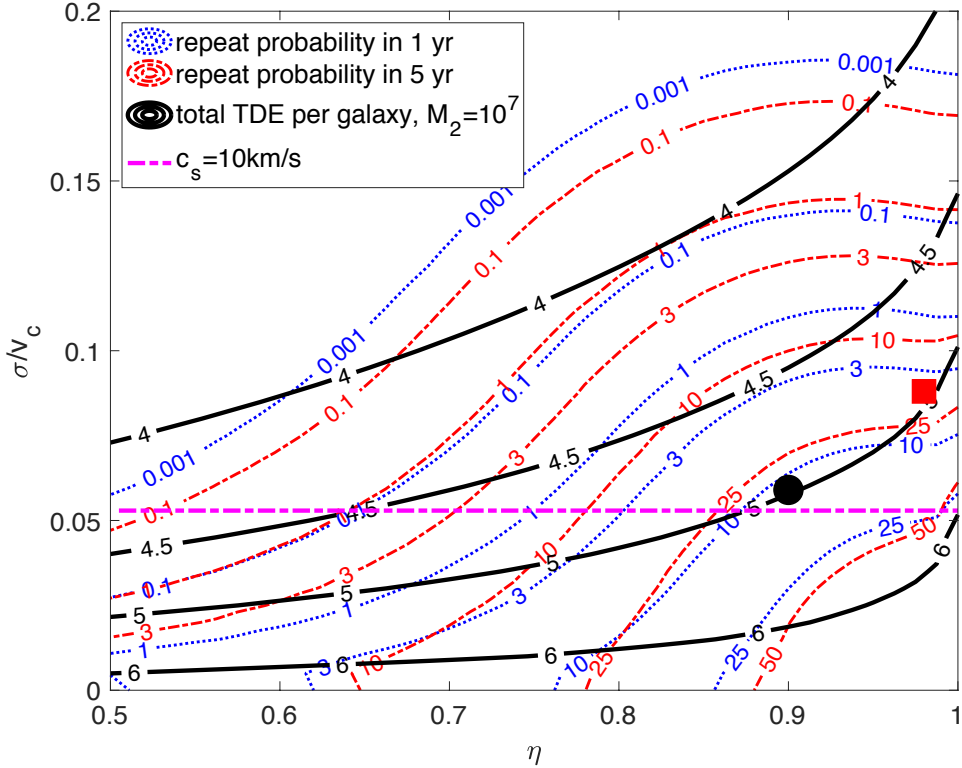}}
\vskip 0.0cm
\caption{
shows contours of probability $P(\Delta t)$ in percent for repeating TDEs within a time interval of $\Delta t=1$yr (blue dotted contours)
and $\Delta t=5$yr (red dot-dashed contours), respectively, on the two-dimensional parameter plane of $(\eta, \sigma/v_c)$,
where $\sigma$ is the velocity dispersion of stars in the disk and $\eta$ is the fraction of stellar mass on the disk.
The black contours are $\log N_{tde}$ per PSB.
Also shown as horizontal magenta dot-dashed line is an indicative case where the vertical velocity dispersion 
is equal to sound speed of atomic cooling gas gas of temperature $10^4$K, 
out of which stars in the disk may have formed.
The fiducial values used are $M_2=10^7\msun$ and $r_{in}=1.2$pc.
Note that in computing the cross section of TDEs, we remove the area inside 
the event horizon of the secondary assuming a Schwarzschild black hole for all calculations.
}
\label{fig:p0}
\end{figure}

In Figure~\ref{fig:p0} 
we place contours of $P(\Delta t)$ for $\Delta t=1$yr (blue dotted contours)
and $\Delta t=5$yr (red dot-dashed contours), respectively,
on the two-dimensional parameter plane of $(\eta, \sigma/v_c)$.
The two black contours are the current observational constraint of $N_{tde}=10^5-10^6$ per PSB, 
corresponding to $10^{-4}-10^{-3}$yr$^{-1}$ per PSB with a time span of $1$Gyr \citep[][]{2016French}
[also see \citet[][]{2017LawSmith, 2018Graur}].
Also shown as horizontal magenta dot-dashed line is an indicative case where the vertical velocity dispersion 
is equal to sound speed of atomic cooling gas gas of temperature $10^4$K, 
out of which stars in the disk may have formed.

Several points are noted.
First, as expected,
the total number of TDEs per PSB tends to increase towards the lower-right corner
of high $\eta$ and low $\sigma/v_c$, due primarily to the increase of the number density of stars in the disk.
Second, if disk thickness is not less than $10\kms$, due to either fragmentation of gas disk at atomic cooling temperature
and/or possible additional heating subsequent to formation of the stellar disk including heating by the secondary itself during
its inspiral, then, an observational constraint of $N_{tde}>10^5$ per PSB
would require $\eta\ge 0.87$ (where the purple line intersects that black contour curve), 
i.e., the disk component is dominant in the inner region. 
Third, an observational constraint of $N_{tde}>10^5$ per PSB 
also indicates that the thickness of the disk cannot exceed $\sigma/v_c\sim 0.1$, a limiting case when $\eta=1$.
Finally, for $N_{tde}=10^5$ per PSB, 
we see that there are regions where a repeater could occur with 
$2-10\%$ probability within a year per PSB in this particular case.
Within five years, there is parameter space where $12-28\%$ probability is seen in this particular case.

\begin{figure}[!h]
\centering
\vskip -0.0cm
\hskip -0.7cm
\resizebox{3.75in}{!}{\includegraphics[angle=0]{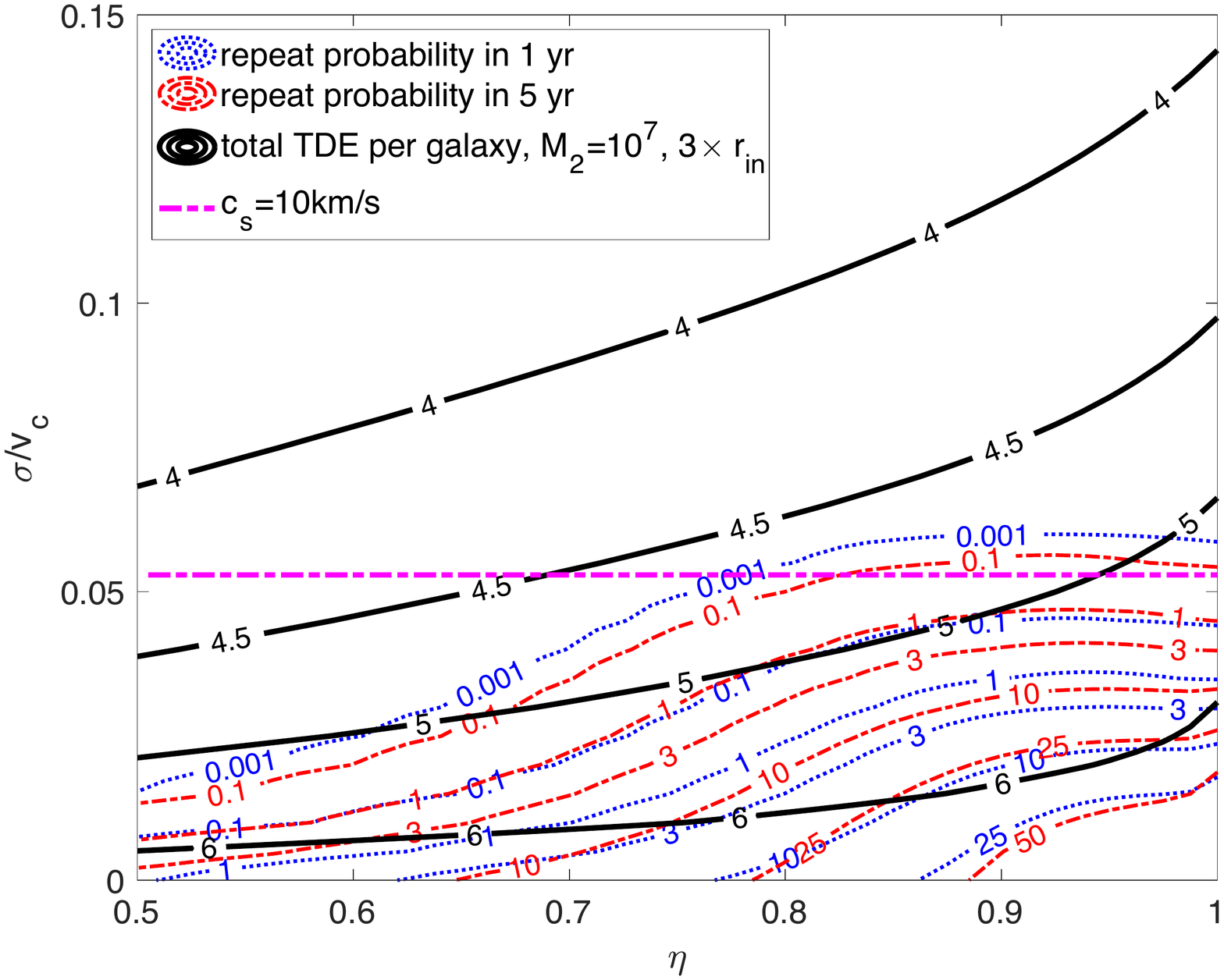}}
\hskip -0.7cm
\resizebox{3.75in}{!}{\includegraphics[angle=0]{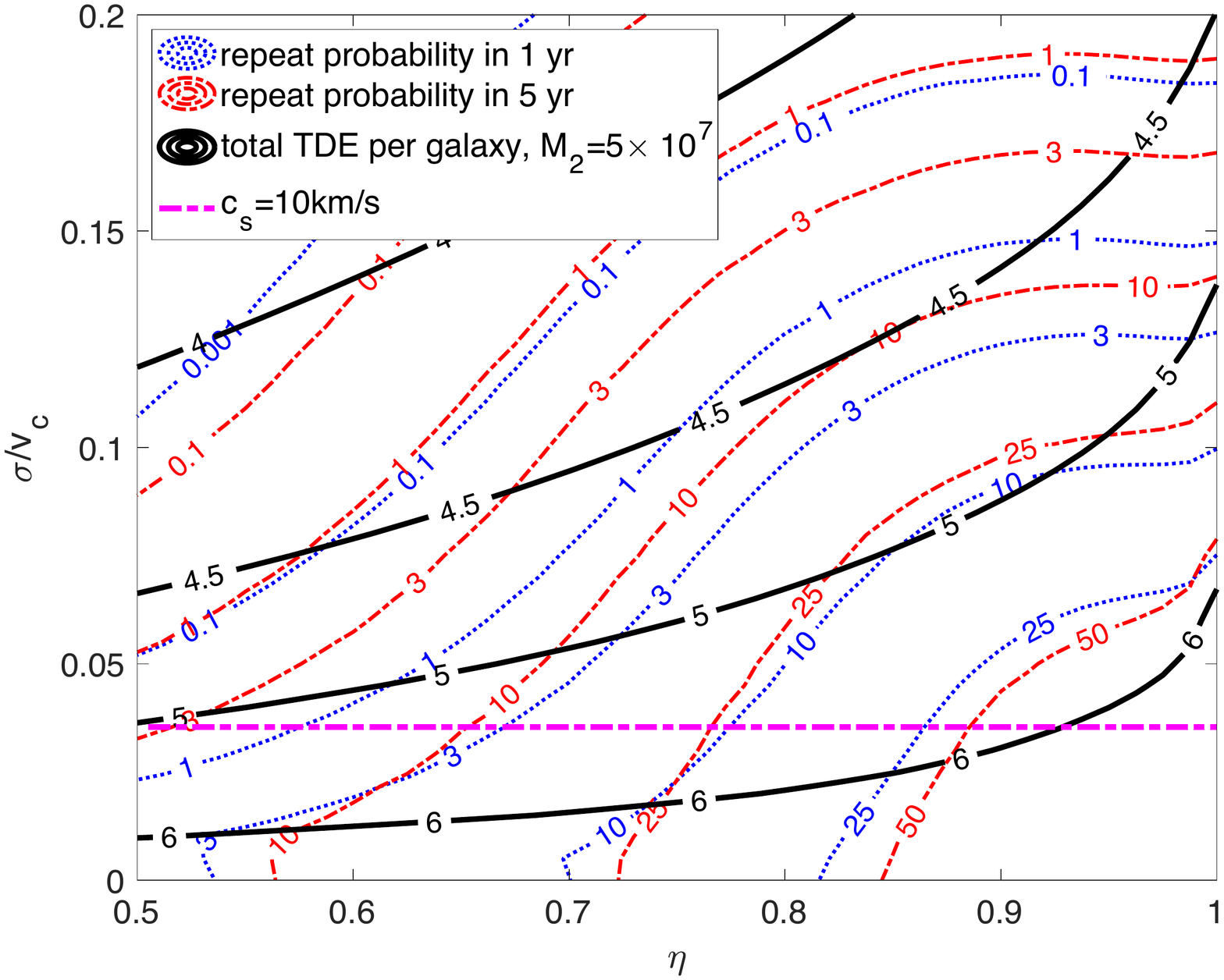}}
\vskip -0.5cm
\hskip -0.7cm
\resizebox{3.75in}{!}{\includegraphics[angle=0]{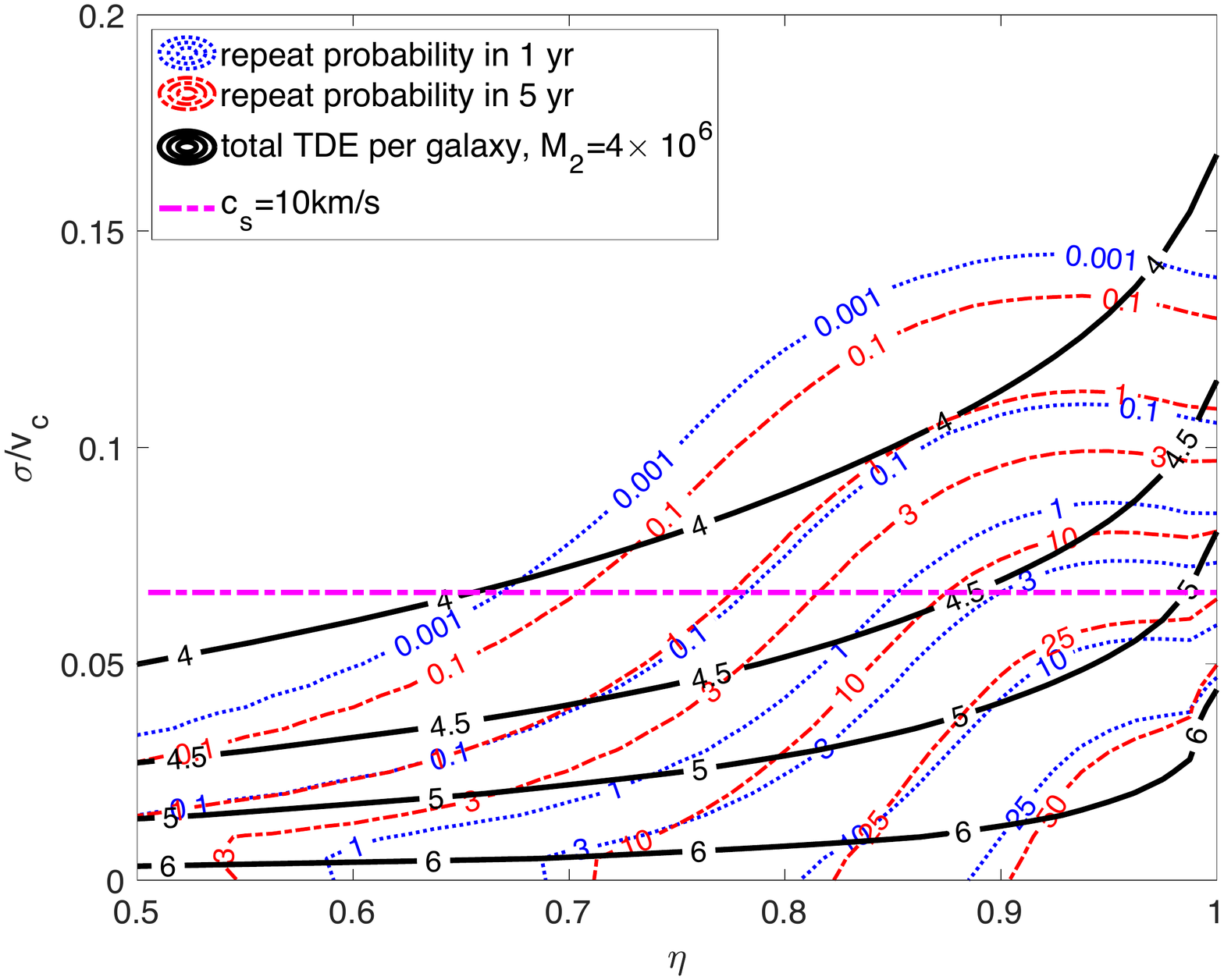}}
\hskip -0.7cm
\resizebox{3.75in}{!}{\includegraphics[angle=0]{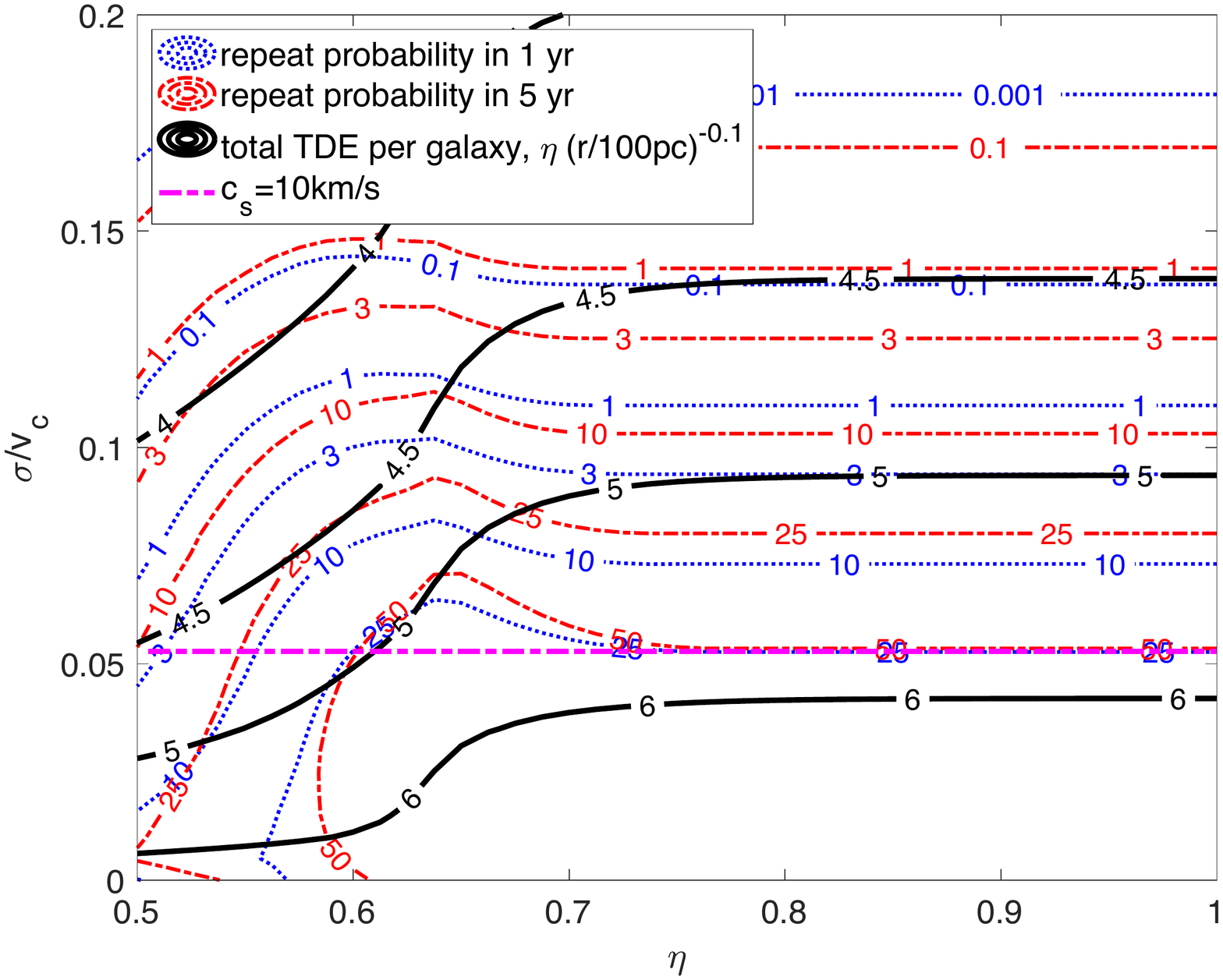}}
\vskip 0.0cm
\caption{
Top-left: the physical parameters of this model are identical to those used for Figure~\ref{fig:p0} except one difference: $r_{in}=3.6$pc instead of $1.2$pc.
Top-right: the physical parameters of this model are identical to those used for 
Figure~\ref{fig:p0} except one difference: $M_2=5\times 10^7\msun$ with an appropriate $r_{in}$ according to Eq (\ref{eq:rin}).
Bottom left: the physical parameters of this model are identical to those used for 
Figure~\ref{fig:p0} except one difference: $M_2=4\times 10^6\msun$ with an appropriate $r_{in}$ according to Eq (\ref{eq:rin}).
Bottom right: the physical parameters of this model are identical to those used for 
Figure~\ref{fig:p0} except one difference: the mass fraction in the disk 
is allowed to increase slowly inward, equal to lesser of $\eta(r/100pc)^{-0.1}$ and unity.
}
\label{fig:p0rest}
\end{figure}

While $r_{in}$ may be low-bounded by Eq (\ref{eq:rin}),
it is possible that star formation may be truncated or flattned at a larger radius.
Thus, we check how results depend on this.
In the top-left panel of Figure~\ref{fig:p0rest} show a case that is the same as that shown in  
Figure~\ref{fig:p0} except one difference: $r_{in}=3.6$pc instead of $1.2$pc.
It is seen that the 
available parameter space for producing 
$N_{tde}=10^5-10^6$ per PSB is compressed towards lower $\sigma/v_c$ and higher $\eta$.  
But there is parameter space still available for explaining the observed abundance of TDEs even in this case.
A large change is in TDE repeat frequencies: it is seen that
there is no parameter space where a TDE may repeat at a probability greater than $0.1\%$ within one year.
There is a limited region in the parameter space where
$0.1\%$ probability exists for a TDE to repeat within $5~$yr.
It is possible to argue both ways as to which physical configuration of the two cases shown is more fine-tuned.
Absence of some introduced scale, it seems more natural to suppose that
the stellar disk could extend to some small radii of no particular choice,
with $r_{in}$ imposed only because of the dynamical reason for the secondary to inspiral, as in 
Eq (\ref{eq:rin}).
Thus, we suggest that $r_{in}=1.2$pc in this case is a less fine-tuned outcome.

Recall that the maximum black hole mass for disrupting a main sequence star is about $10^8\msun$ for a Schwarzschild black hole
(see Eq \ref{eq:rin}).
The top-right panel of Figure~\ref{fig:p0rest} displays the case for $M_2=5\times 10^7\msun$ with an appropriate $r_{in}$ according to Eq (\ref{eq:rin}).
A comparison between it and Figure~\ref{fig:p0} 
indicates that a more massive black hole tends to only slightly enhance both the overall rate of TDEs per galaxy 
and the probability of repeaters on relevant times scales.
However, the range of $\eta$ for achieving the same $N_{tde}=10^5$ is enlarged,
when constraining $\sigma\sim 10\kms$, to $\eta\ge 0.5$.
But the overall rate and repeater probability contours do not change dramatically.
The reason for this week dependence on $M_2$ is due to a larger, removed cross section inside the event horizon that
almost compensates the increased tidal radius for a larger black hole,
among other factors.

Next, we consider a case of merger of two lower mass galaxies, with $M_1=M_2=4\times 10^6\msun$,
and $r_{in}$ determined according to Eq (\ref{eq:rin}).
The bottom-left panel of Figure~\ref{fig:p0rest} shows the result, for which we note three points.
First, the model can no longer accommodate the observed $>10^5$ TDEs per PSB,
except in a very small parameter space at $\eta>0.98$ and $\sigma/v_c\sim 0.07-0.08$.
Second, in the available parameter space,
the repeating rate is, however, comparable to the fiducial case shown in Figure~\ref{fig:p0}. 
Combining results for the models,
we conclude that, while  
the overall abundance of TDEs increases with the black hole mass,
the repeating rate per PSB depend weakly on the SMBH mass for a given $N_{tde}$,
as long as the inner radius of the central stellar disk is not cutoff.

Finally, 
the bottom-right panel of Figure~\ref{fig:p0rest} shows the result 
for a case where we let the mass fraction of the disk component to 
increase inward from $100$pc to $r_{in}$ as $\eta (r/100)^{-0.1}$, capped of course at unity.
We see that the available parameter space is significantly enlarged compared to the fiducial case,
with the shape of contours seen to flatten out horizontally,
while the repeater probability at a given $N_{tde}$ remains roughly in the same range.

To summarize, in our model the overall rate of TDEs per PSB, averaged over time,
is set by the long dynamical friction process for the secondary to inspiral following galaxy merger.
A unique characteristics of our model is that  
once having reached and aligned with the central stellar disk,
the overall migration time interval over which the bulk of the TDEs 
occur is much shorter than the typical lifetime of PSBs of $\sim 1$Gyr.
Consequently, one important prediction of this model is that
TDEs may repeat on a reasonable time scale.
While a precise repeating rate is difficult to nail down, 
because we are not certain about the parameter space of $(\eta, \sigma/v_c)$ that nature picks,
we see that within (1,5) years the repeating probability
falls in the range of ($0.1-10\%$, $3-30\%$)
 if $N_{tde}=10^5$,
under the condition that $M_2=4\times 10^6-5\times 10^7\msun$ and no inner cuttoff of stellar disk 
(i.e., $r_{in}$ is determined by Eq \ref{eq:rin}).
Thus, assuming $N_{tde}=10^5$ and with a sample of $1000$ TDEs,
it appears that at least one repeat may be detected within one year;
alternatively, with a sample of $30$ TDEs,
at least one repeat may be detected within five years.
If the current observationally inferred $N_{tde}$ range of $10^5-10^6$ indeed holds up,
the above estimated range of repeating probability would be an under-estimate.
If observations do find such repeaters, they would provide strong support for this model.
With enough statistics and time baseline, it may then be possible to tease out useful information 
on the physical configuration of the central disk
in terms of parameter space of $(\eta, \sigma/v_c, M_2)$.
A statistical comparison between the number of PSBs with TDEs and those without 
may additionally shed luminous light on the temporal distribution of TDEs in PSBs and the distribution of the time 
for the secondary to land on the central stellar disk, which may be ultimately linked to galaxy formation process.
As a reference, in a model with a delay time distribution (DTD) 
of $t^{-0.5}$ \citep[][]{2018Stone}, generously extending to $1$Myr at low end and normalized to $N_{tde}=10^6$ TDEs over $1$Gyr in a PSB,
the probability of repeaters within five years is practically zero ($2.1\times 10^{-32}$).

It is appropriate to prudently ask the following question:
is the condition that required to accommodate the observed TDE rates in PSBs physically plausible?
In particular, is $\sigma/v_c\sim 0.1$ viable?
Let us examine what this means with respect to 
the column density, volumetric density and temperature of the gas disk forming the disk stars.
Adopting $v_c=189\kms$ for a Mestel disk,
we find that surface density ${\rm \Sigma(r) = 276 (r/1pc)^{-1} g~cm^{-2}}$ and
a volumetric density ${\rm n_H=2.7\times 10^8 (v/10\sigma)(r/1pc)^{-2}~cm^{-3}}$.
The mid-plane pressure due to gravitational mass above  
is 
${\rm p=\pi G \Sigma^2(r)/2 = 8.0\times 10^{-4} (r/1pc)^{-2} dyn~cm^{-2}}$.
This means that, if the downward gravity is balanced by thermal pressure,
the gas temperature would have to be $2.2\times 10^4 (10\sigma/v_c)~K$,
where a molecular weight of unity is used for simplicity.
We see that 
$\sigma/v_c = 0.1$ and $0.05$
would imply a gas temperature of $2.2\times 10^4$~K and $1.1\times 10^4$~K,
respectively.
This is in the exact regime where gas has been cooled rapidly by atomic cooling processes after infall shock
but has yet to be cooled further down by molecular cooling (and low temperature metal cooling) processes.
At a density of $5.4\times 10^8~cm^{-3}$ and ${\rm T=1.1\times 10^4}$K for $\sigma/v_c=0.05$ at $r=1$pc,
the Jeans mass is $7.3\times 10^2\msun$.
It indicates that the gas disk at $r\sim 1$pc would fragment at $T\sim 10^4$K, which may subsequently
form stars directly from atomic cooling gas or may go through the molecular phase first and then form stars.
In either case, it appears quite plausible that a disk of 
height to radius ratio of $0.05-0.1$ for $v_c=189\kms$ at $r\sim 1$pc and larger radii
(note the weaker increase of Jeans mass than the mass on the Mestel disk with increasing radius at a given gas temperature).
It is in fact quite remarkable that this completely independent assessment of
the likely $\sigma/v_c$ from a physical point of view of gas cooling and fragmentation 
is almost exactly what is required for producing the observed abundance of TDEs in PSBs.

\subsection{TDEs Spatially Offset from Center and Complexities of Debris Dynamics}

\begin{figure}[!h]
\centering
\vskip -0.0cm
\resizebox{6.0in}{!}{\includegraphics[angle=0]{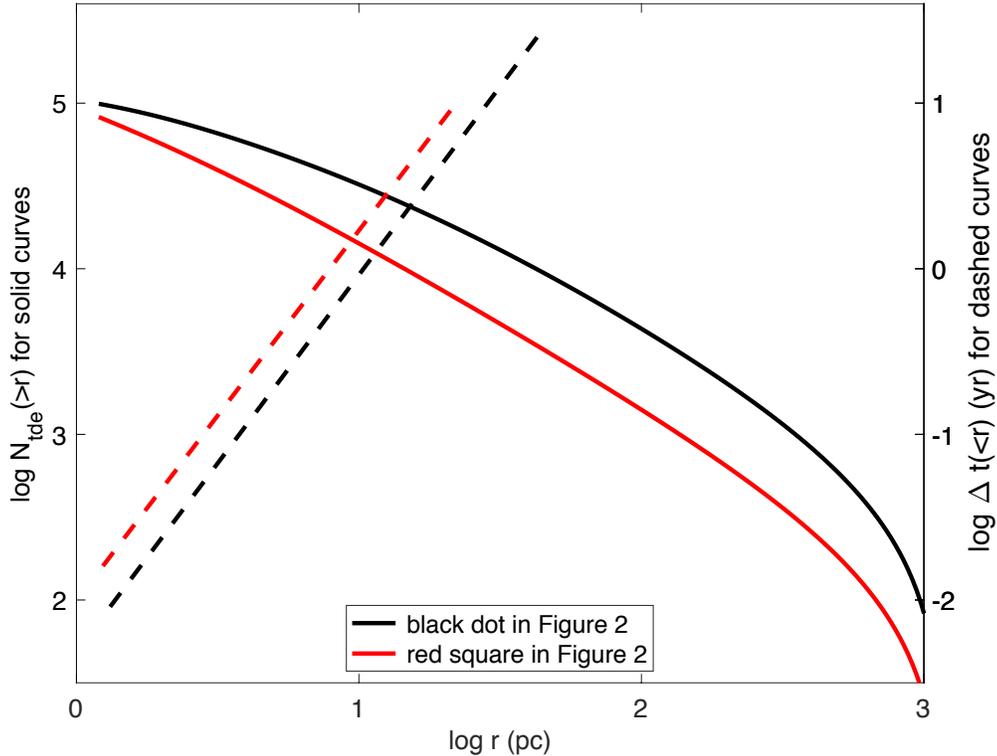}}
\vskip 0.0cm
\caption{
shows as solid curves the cumulative (from large to small radius) radial distributions of TDEs for two cases,
corresponding to the case with $(\eta,\sigma/v_c)=(0.9,0.059)$ shown as a solid black dot in Figure~\ref{fig:p0}
and the case with $(\eta,\sigma/v_c)=(0.98,0.088)$ shown as a solid red square in Figure~\ref{fig:p0}.
Also shown as corresponding dashed curves 
are the mean interval between successive TDEs cumulative up to that point from the small radius end,
as indicated by the right y-axis.
}
\label{fig:Nr}
\end{figure}

TDEs in our model do not occur about the central black hole.
Figure~\ref{fig:Nr} 
shows the cumulative (from large to small radius) radial distributions of TDEs for two
combinations of $(\eta,\sigma/v_c)$ indicated 
as the black dot and red square in Figure~\ref{fig:p0}.
We see that depending on the physical configuration ($\eta$, $\sigma$, $v_c$, $r_{in}$ and $M_2$) 
of the system the distribution varies.
Despite the uncertainties, in the two cases it is seen that 
$90\%$ of TDEs occur with a spatial offset of $15-50$pc from the central black hole.
Trying to detect this spatial offset will be a challenging but rewarding undertaking.
If confirmed, it would provide unambiguous evidence for TDEs occurring around the secondary that is inspiraling
and would have profound implications for dynamics of SMBH inspiral and galaxy formation as a whole.
It will also have important implications for future gravitational wave observations aimed at detecting
mergers of SMBHs, e.g., by LISA.
As an example, an angular separation of $0^".1$ would correspond to an offset of $12$pc at a distance of $25$Mpc.
Such an offset would have been resolved even with current capabilities.
\citet[][]{2016Leloudas} measure a spatial offset of $131\pm 192$pc, corresponding to an
angular offset of ${\rm 36~mas \pm 53~mas}$ for ASASSN-15lh at $z=0.2326$;
in this case, due to the large distance the offset is ambiguous.
Combining detections of TDEs from large upcoming surveys,
including the Zwicky Transient Facility (ZTF),
eROSITA, the Large Synoptic Survey Telescope (LSST),
with the next-generation large ground-based telescopes with AO capabilities may routinely 
enable this for nearby sources, which will certainly provide a strong test of the model once a sufficient sample is produced and
statistical characterization possible.

For a typical main sequence star TDE, the inbound debris at the deepest orbits has an energy of 
about $(-v_{orb}v_*)$, where $v_{orb}$ is orbital velocity at the tidal radius and $v_*$ stellar escape velocity,
roughly corresponding to an orbital period of order one year.
Debris with an energy that is ten times lower will have an orbital period of about $30$yr, and so on.
Let us take debris on a one-year orbit to illustrate.
For a secondary of mass $10^7\msun$, it corresponds to an eccentric orbit with a semi-major axis of $\sim 10^{-3}$pc.
If the period of the secondary is on an orbit of radius $1$pc about the center of the galaxy, then its period is $3.3\times 10^4$yr
and the rate of spatial displacement of the secondary would be $\sim 10^{-4}$pc/yr. 
The tidal radius of main sequence stars about the secondary is $5\times 10^{-6}$pc (Eq \ref{eq:rt}).
Thus, when the debris is at the apobothron about the secondary,
the effect of accelerated motion of the secondary is at a level of $10^{-4}pc/10^{-3}pc\sim 10\%$ per orbit of the debris,
where the debris may be be able to make orbital adjustments to follow the gravitational center of the secondary
in a fashion akin to a three-dimension precession.
However, when the debris is at the peribothron (i.e., around the tidal radius),
it will not be able to make orbital adjustments promptly enough to form a smooth orbit due to the accelerated
motion of the secondary.
While detailed simulations will be required to obtain more quantitatively precise results
for a wide variety of orbital configurations of the secondary,
it is not difficult to see that light curves and also the spectral properties 
of TDEs could be much more varied than the case of a static black hole.

The static SMBH at the center of the galaxy may also produce TDEs.
In this case, the relativistic precession is often invoked to cause debris to interact and circularize.
For reference, we merely restate 
that the GR precession period $P_{pr}$ of a debris with an orbital period $P_{de}$ with eccentricity $e$ at the tidal radius is 
\begin{equation}
\label{eq:t}
  \begin{split}
{\rm P_{pr} = {P_{de} (1-e^2)\over 3} ({c\over v_t})^2 = 3.4 P_{de} (1-e^2) ({M_2\over 10^7\msun})^{-2/3} ({r_*\over \rsun}) ({m_*\over \msun})^{-/3}},
  \end{split}
\end{equation}
\noindent
where $v_t$ is the orbital velocity at tidal radius as the peribothron.
It is seen that for debris on orbits of periods on the order of one year,
relativistic precession will circularize debris on the order of the period.
In this simpler case, the structure would remain two-dimensional throughout.
Therefore, light curves are expected to be smooth.

\subsection{Narrow and Extended Narrow Line Regions in Post-Starburst Galaxies}

Are there some possible effects of the UV radiation from the TDEs on gas on galactic scales?
Let us examine three time scales.
First, the light travel time to $1-10$kpc from the central region where TDEs take place
is about $t_{tr} = 3\times 10^3 - 3\times 10^4$yr.
Second, for a gas cloud of hydrogen volume density $n_H$ and temperature of $10^4$K, the
recombination time is $t_{rec}=1.2\times 10^3 (n_H/100cm^{-3})^{-1}$yr.
Third, the time interval between successive TDE events, $\Delta t_{tde}$.
In Figure~\ref{fig:Nr} we show as dashed lines $\Delta t_{tde}$ cumulatively as a function of radius inside-out.
We see $\Delta t_{tde}\sim 1-30$yr over the bulk of the $\sim 10^5$ TDE events in each case.
If both conditions, $\Delta t_{tde} \le t_{tr}$ and $\Delta t_{tde} \le t_{rec}$,
 are satisfied, which is true as long as $n_H$ is not greater than $10^4cm^{-3}$ or so,
then, the gas cloud would ``see" a steady UV light bulb at the center over a period of $1$Myr or so, and is kept fully ionized 
despite the actual ``flickering" of light, over this period of a flurry of TDEs.
While the electron densities in some inner narrow line regions may exceed ${\rm 10^4~cm^{-3}}$,
the observationally inferred electron densities 
on scales of kpc or larger 
are found to be in the range from a few ${\rm 10^2~cm^{-3}}$ to mid ${\rm 10^3~cm^{-3}}$
\citep[e.g.,][]{2006Nesvadba, 2011Greene, 2013Arav}.

One notes that for a given PSB of an age $0.1-1$Gyr, this intense period of TDEs makes up
only small fraction, $0.1-10\%$, of the their overall age.
Thus, it is likely that a blind survey of PSBs will only see a small fraction of them
capable of enabling narrow or extended narrow line regions. 
For PSBs with observed TDEs it may be expected that 
most of them may possess narrow or extended narrow line regions,
to the extent that suitable gas clouds on these scales exists.
In other words, the fraction of PSBs with narrow or extended narrow line regions 
for those experiencing TDEs is now primarily a function of the abundance of gas clouds
on these scales, not the availability of radiation sources to ionize them.
Since PSBs have experienced a significant merger event in the last $0.1-1$Gyrs,
it may be that such gas clouds are ubiquitous due to the merger
and/or subsequent starburst driven outflows and/or interactions between outflows and accreting gas on galactic scales.
We suggest that an observational campaign to search for narrow or extended narrow line regions
in PSBs, especially those with detected TDEs,
may be very useful to provide an additional, indirect test of the model.
Possibly related, the observed enhancement of LINERs in PSB galaxies 
\citep[e.g.,][]{2006Yan, 2006Yang}
may be powered by TDEs in both radiation and, in part, the emitting gas that is the unbound debris.

\section{Conclusions}

Under a simple tractable example of an isothermal stellar distribution 
with a Mestel disk component of mass fraction $\eta$ with 
a small velocity dispersion 
($\sigma/v_c \le 0.1$) 
and a three-dimensional spherical component of mass fraction $1-\eta$,
we calculate the dynamics of the inspiral of the secondary SMBH through the disk midplane,
with the aim to quantify the tidal disruption event rates by the inspiraling SMBH.
We envision that such a central stellar disk of size $\le 100$pc may be produced following a galaxy merger 
and it a part of the starburst that subsequently fades to turn into a post-starburst galaxy some $0.1-1$Gyr later.
Insofaras the secondary is made co-planar with the disk,
the results depend little on either the outer or the inner radius of the disk,
as long as the disk is not cutoff prematurely in the interior.
Coplanarity may be inevitable for a system even with a moderate amount of oblateness and an anisotropic velocity distribution,
where the dynamical friction drag tends to align the inspiraling object with the disk plane on the dynamical friction time scale. 
At radii larger than $30-100$pc (depending on the relative mass fractions of the disk and the bulge components)
the inspiraling SMBH moves relative to the stars in the disk, at a low velocity often referred to the asymmetric drift in our Galaxy, 
and experiences an efficient two-dimensional dynamical frictional force.
At smaller radii, dynamical friction of the secondary with the bulge stars may be expected to be more efficient
in the viable parameter space in our model. 
In either regime, the TDEs are produced by stars in the disk.

The two main parameters are the ratio of velocity dispersion to the circular velocity $\sigma/v_c$ and
the fraction mass of the disk component $\eta$ in the region.
With $\sigma/v_c\sim 0.03-0.15$ and $\eta>0.7$,
it is shown that $10^5-10^6$ TDEs of solar mass main sequence stars 
per post-starburst galaxy may be produced by the secondary of mass $10^6-10^8\msun$.
The vast majority of TDEs by the secondary in a post-starbust galaxy occurs within a space of time of $\sim 30$Myr or shorter.
Thus, the apparent age distribution of post-starburst galaxies of $\sim 0.1-1$Gyr 
is not the duration over which TDEs occur in any individual post-starburst galaxy, 
rather it reflects the rich variety of galaxy mergers 
and the range in time that it takes to bring the secondary to the central disk.
To further test this model, we provide five unique predictions.

$\bullet$ 
A unique prediction of this model is that TDEs may repeat on a time scale amenable to astronomers.
While a precise repeating rate is difficult to nail down,
our model shows that, normalizing to a rate of $10^5$ TDEs per post-starburst galaxy,
with a sample of $1000$ TDEs, at least one repeater may be detected within one year.
Alternatively, on a time scale of five years with a sample of $30$ TDEs,
at least one repeater may be expected to occur.
This will be imminently testable.

$\bullet$ 
The second unique prediction is that the TDEs are expected to display a spatial offset from the galactic center of $\sim 1-300$pc for $\ge 99\%$ of the TDEs.
Combining upcoming detections of TDEs from ZTF, eROSITA and LSST
with the next-generation large ground-based telescopes with AO capabilities 
should be able to detect this within a distance of $50-500$Mpc for offsets of $1-10$pc, which will provide an unambiguous test of the model,
once a sufficient sample is produced and statistical characterization made.
If detected, it will also shed useful light on future gravitational wave observations aimed at detecting mergers of SMBHs, e.g., by LISA.

$\bullet$ 
Third, the accelerated motion of the secondary during the return flight of the debris may cause their orbits 
to be three-dimensional to form a three-dimensional structure at some radius
outside the tidal radius.
This may complicate the prediction and possibly leads to a rich variety of light curves of TDEs.
Detailed calculation on this front is deferred.

$\bullet$ 
Fourth, the high cadence of TDEs may serve as a quasi-continuous UV source for galactic scale gas,
since both the light travel and gas recombination times are longer that the mean cadence.
Thus, it may be expected that post-starburst galaxies, especially those with detected TDEs,
may possess narrow or extended narrow line regions, to the extent that the galaxy merger and/or the starburst driven outflows
have created suitable clouds there for the radiation to illuminate.  A systematic survey can verify this. 

$\bullet$ 
Finally, since it is the inspiraling secondary disrupting the stars,
the central SMBH is no longer required to be less massive than $10^8\msun$ for a galaxy to produce TDEs.
We note that the mass of the central SMBH for the TDE event ASASSN-15lh \citep[][]{2016Leloudas} 
is inferred to be $(3-6)\times 10^8\msun$, based on the stellar mass or luminosity of the host galaxy,
which would solidly place it in the impotent SMBH camp for producing TDEs of main sequence stars.
The authors suggest a Kerr black hole to marginally get by.
In our model, this is not a problem, since the mass of the secondary black hole could be lower than that of the central one.
Incidentally, this TDE has a measured spatial offset of $131$pc from the galactic center, albeit with an undesirable $1\sigma$ errorbar of $192$pc
presently.

\vskip 1cm

I would like to thank an anonymous referee for critical and constructive reports
and for checking every single term of the equations that greatly helped improve the paper.
I would like to thank Ben Shappe, Decker French, Jane Dai, Iair Acavi and Tsvi Piran for helpful discussion,
and Nick Stone for a wonderful talk that spawned this inquiry.
I would like to thank Yukawa Insitute of Theoretical Physics for providing opportunity at short notice
in the Yukawa seminar to give a talk on this work pre-publication.
The research is supported in part by NASA grant 80NSSC18K1101.


\end{document}